\documentclass[conference]{IEEEtran}
\IEEEoverridecommandlockouts
\usepackage{cite}
\usepackage{amsmath,amssymb,amsfonts}
\usepackage{algorithmic}
\usepackage{graphicx}
\usepackage{textcomp}
\usepackage{tabularx}

\usepackage{tabularx}
\usepackage{array}
\usepackage{booktabs}
\usepackage{xcolor}
\usepackage{graphicx}
\usepackage{cite}
\usepackage{picinpar}
\usepackage{amsmath}
\usepackage{url}
\usepackage{flushend}
\usepackage[utf8]{inputenc}
\usepackage{colortbl}
\usepackage{soul}
\usepackage{multirow}
\usepackage{makecell}
\usepackage{pifont}
\usepackage{xcolor}
\usepackage{alltt}
\usepackage[hidelinks]{hyperref}
\usepackage{enumerate}
\usepackage{siunitx}
\usepackage{breakurl}
\usepackage{epstopdf}
\usepackage{pbox}
\usepackage{amsfonts}
\usepackage{mathtools}
\usepackage{commath}
\usepackage{booktabs}
\usepackage{caption}
\usepackage{subcaption}
\usepackage{bm}
\usepackage{mathrsfs}
\usepackage{amsthm}
\theoremstyle{remark}

\usepackage[linesnumbered,ruled,vlined]{algorithm2e}

\def\BibTeX{{\rm B\kern-.05em{\sc i\kern-.025em b}\kern-.08em
    T\kern-.1667em\lower.7ex\hbox{E}\kern-.125emX}}

\begin{document}

\title{Online Learning-Based Adaptive Hybrid Benders Decomposition for Risk-Averse Optimal Sizing 
\thanks{This research was partially funded by the following projects: (a.) CETPartnership, the Clean Energy Transition Partnership under the 2023 joint call for research proposals, co-funded by the European Commission (GA N°101069750) and with the funding organizations detailed on https://cetpartnership.eu/funding-agencies-and-call-modules and (b.) ‘Flexibility services based on Connected and interoperable Hybrid Energy Storage System (FlexCHESS)’ funded by the European Union under the H2020 Innovation Framework Programme, GA No 101096946\\
$^{\text{a}}$ Affiliated with Aix Marseille Universit\'e, LIS Lab CNRS UMR 7020, Campus Saint-J\'er\^ome, 13013 Marseille, France.\\
$^{\text{b}}$ Affiliated with the School of Electrical and Electronic Engineering, The University of Sheffield, S1 3JD Sheffield, U.K. and Autonex Systems Limited, Coventry CV1 2NT, U.K.\\
\texttt{email: saif.ahmad,seifeddine.benelghali@\{lis-lab.fr\}, hafiz.h.ahmed@ieee.org}
}
}

\author{\IEEEauthorblockN{Saif Ahmad$^{\text{a}}$}
\and
\IEEEauthorblockN{Seifeddine BEN ELGHALI$^{\text{a}}$}
\and
\IEEEauthorblockN{Hafiz Ahmed$^{\text{b}}$}
}

\maketitle
\IEEEpubidadjcol
\begin{abstract}
Optimal sizing problem (OSP) for battery energy storage system (BESS) under uncertain inputs is often formulated as a two-stage stochastic program (2SP), which typically introduces computational and RAM (memory) bottlenecks for large scenario sets. Benders Decomposition (BD) is a popular approach for tackling this problem, but it suffers from slow convergence to the exact solution. To address this problem, we propose an accelerated online learning-based adaptive hybrid BD algorithm for a risk-averse 2SP formulation. The proposed method avoids getting stuck in the infeasible region during early iterations by explicitly embedding a carefully selected scenario in the master problem, while a tail-relevant scenario selector based on online learning helps to avoid solving the entire scenario set at every iteration. The OSP is formulated to select a behind-the-meter BESS in a multi-site energy community with existing renewables. The use case explores an interesting middle-ground between deterministic acceleration methods and training-heavy ML models, showcasing the potential of ML-assisted decision-making. Compared with vanilla BD, OLAH-BD reduces
subproblem evaluations and total wall time by up to
 80\% under the same tolerance settings.

\end{abstract}


\section{Introduction}
Decarbonization of the urban energy mix through smart aggregation, such as energy communities (ECs), is a promising concept that prioritizes collective self-consumption of renewables and can alleviate stress on the distribution network during peak times. This work focuses on the economic sizing of battery energy storage systems (BESS) in ECs with preinstalled PV on one or more sites, a recurring scenario driven by declining PV module costs and high feed-in tariffs that spurred initial investments. However, recent regulatory changes nudge operators towards BESS installation to remain profitable.  

Optimal BESS sizing under uncertainty in renewables, load, etc., can often be formulated as a two-stage stochastic program (2SP), which involves making first-stage (here-and-now) decisions before uncertainty is realized, followed by second-stage (recourse) decisions under the assumption of perfect scenario foresight \cite{patel2022neur2sp}. We consider an illustrative example for sizing a behind-the-meter BESS in energy communities ECs having multiple sites that can exchange energy via the local distribution network. In most cases, 2SPs are solved by approximating the uncertainty with a discrete set of possible realizations, called scenarios. A monolithic 2SP formulation with copies of the recourse costs + constraints per scenario becomes computationally intractable as the scenario size grows \cite{patel2022neur2sp}. In the BD framework, a large monolithic 2SP model is split into a master problem containing first-stage costs and subproblems corresponding to scenarios with fixed first-stage variables, which can be solved independently in parallel and in batches \cite{benders1962partitioning}. Although the model split helps with computation (and memory), solving the master problem and then the subproblems (SPs) iteratively to reach convergence drastically slows the process \cite{rahmaniani2017benders}. 

In recent years, much effort has been devoted to developing acceleration methods that reduce the computation time for BD, such as those targeting the relaxed master problem (RMP) \cite{jia2021LearnDB,cai2026learning2cut,ramirez2023bendersadaptivecuts}, while others aim to speed up the SP evaluation or cut generation \cite{guan2026proxyBD,blanchot2023bendersbybatch,donkiewicz2026adaptiveBDlogisticregression,magnanti1981accelerating}. Managing RMP growth via cut filtering \cite{cai2026learning2cut,jia2021LearnDB} or aggregation \cite{ramirez2023bendersadaptivecuts} is an effective approach that reduces total run time, especially when the RMP is computationally heavy. On the other hand, batch-wise scenario selection in each iteration \cite{blanchot2023bendersbybatch,ramirez2023bendersadaptivecuts} and modeling the recourse problem/cut coefficients using neural networks (NNs) \cite{liu2025icnn,patel2022neur2sp,guan2026proxyBD} are powerful methods to reduce scenario computation time. 
Stochastic and random scenario selection-based BD algorithms with verified convergence guarantees have been proposed in \cite{bertsimas2025stochasticBD,pauphilet2025surprisingRandomPartialBD}. However, random scenario selection may result in worse performance than a deterministic baseline \cite{donkiewicz2026adaptiveBDlogisticregression}. A comprehensive review of the acceleration methods used in BD framework is presented in \cite{rahmaniani2017benders}.  Although the modern version of BD is usually implemented via a callback method, made possible by the RMP having integer variables, a bulk of these approaches readily translate to the more conventional sequential iterative solve. We follow the latter implementation in this work as our model is simplified to be an LP. 

We introduce an online learning-based adaptive hybrid BD (OLAH-BD) algorithm for solving a risk-averse OSP formulated as a 2SP. Our contribution is threefold: \textbf{a.)} we propose a hybrid RMP as a base where one high-stress scenario is embedded explicitly, while the remaining scenarios are solved batch-wise in parallel on CPU threads \textbf{b.)} a regime-based online scenario selector where the initial full sweeps are for stabilization and training a logistic regression (LR) model, and then switches to exploitation with periodic validation via full sweeps and re-learning the weights when required, and \textbf{c.)} optimality-cut filtering such that tail-relevant scenarios are prioritized. Periodic full sweeps recover missed information and provide exact convergence certificates while additionally allowing online retraining of the LR model on a more relevant dataset when required.

\textit{Notation:}
For a finite set $\mathcal A$, $|\mathcal A|$ denotes its cardinality,
$\mathcal A\setminus\mathcal B$ the set difference, and
$\mathcal A\cup\mathcal B$ and $\mathcal A\cap\mathcal B$ the union and
intersection, respectively. The operators
$[z]_+=\max\{z,0\}$ and $\lceil z\rceil$ denote the positive part and
ceiling of $z$. The operator $\arg\max_{s\in\mathcal S}(\cdot)$ returns
an index attaining the maximum, while
$\operatorname{Top}_{m}\{z_s:s\in\mathcal S\}$ denotes the set of
$m$ indices associated with the largest values of $z_s$.

\section{Problem Formulation} \label{sec:problem formulation}
In contrast to a risk-neutral formulation that gives equal importance to each scenario, a risk-averse formulation based on conditional value at risk (CVaR) \cite{rockafellar2000optimization} penalizes the worst outcomes lying in the tail end of the distribution defined by the confidence level $\alpha$. Let $x\in \mathcal X, \ y_s \in \mathcal Y_s$ denote the first- and second-stage variables, respectively, where $s\in S$ denotes the observed scenarios. With the scenario recourse problem defined as
\begin{equation}
\begin{split}
    Q_s(x):=\min_{y_s\in \mathcal Y_s}
    \quad & q_s^\top  y_s\\
     \text{s.t.} \quad & M y_s = m_s+W_sx,\\ 
    & N y_s \le n_s+R_sx,
\end{split}
\label{eq:generic_recourse}
\end{equation}
the CVaR based risk-averse 2SP takes the following form
\begin{subequations}
\begin{align}
\min_{x,\eta,\xi}\quad
& c^\top x+\eta+
\frac{1}{1-\alpha}
\sum_{s\in\mathcal S}\pi_s\xi_s
\label{eq:generic_cvar_obj}\\
\text{s.t.}\quad
& x\in \mathcal X,\\
& \xi_s\ge Q_s(x)-\eta,\qquad \forall s\in\mathcal S,\\
& \xi_s\ge 0,\qquad \forall s\in\mathcal S.
\end{align}
\label{eq:generic_cvar_2sp}
\end{subequations}
where $\pi_s$ is the scenario probability, $\{q_s,m_s,n_s,W_s,R_s\}$ are realisation of the random variables in each scenario while $\{M,N\}$ are considered to be deterministic for most cases. Here $\eta$ is the value-at-risk threshold and $\xi_s$ represents the positive excess cost above $\eta$.

In the numerical study of Section~\ref{sec:numerical},
$Q_s(x)$ represents the optimal annual dispatch cost under PV--load
scenario $s$. The CVaR objective is constructed across the scenario
recourse costs $\{Q_s(x)\}_{s\in\mathcal S}$ through $\eta$ and
$\xi_s$. The first-stage vector $x$ contains BESS sizing and subscription
decisions, while $y_s$ contains the scenario-dependent directed
power-flow variables. The complete application-specific formulation is
given in Appendix~\ref{app:bess_formulation}.

\section{Online Learning-Based Adaptive Hybrid Benders}
\label{sec:method}

We propose an online learning-based adaptive hybrid Benders decomposition
(OLAH-BD) for the 2SP in
\eqref{eq:generic_recourse}--\eqref{eq:generic_cvar_2sp} in this section. Our approach
combines three elements to reduce SP evaluations and accelerate
convergence while retaining full-scenario certification: an explicitly
embedded stress scenario, online logistic-regression (LR)-based scenario
prioritization, and periodic full sweeps with adaptive batch control. 

\subsection{Hybrid Relaxed Master Problem}

In vanilla BD (VBD), the first-stage and CVaR variables form the RMP,
while all scenario-dependent recourse variables are decomposed into SPs.
OLAH-BD instead selects one stressful scenario
\begin{equation}
s^e=\arg\max_{s\in\mathcal S}\psi_s ,
\label{eq:stress_scenario}
\end{equation}
where $\psi_s$ is based on normalized annual deficit energy and peak
deficit, and embeds its complete recourse model in the RMP. The remaining
scenarios form
$
\mathcal D:=\mathcal S\setminus\{s^e\}.
$
Embedding $s^e$ strengthens the RMP by enforcing one complete operational
realization directly, thereby guiding early first-stage decisions away
from strongly infeasible regions. The hybrid RMP is
\begin{subequations}
\label{eq:hybrid_master}
\begin{align}
\min \quad
& c^\top x+\eta+
\frac{1}{1-\alpha}
\left(
\pi_e\xi_e+
\sum_{s\in\mathcal D}\pi_s\xi_s
\right)
\label{eq:hybrid_master_obj}\\
\text{s.t.}\quad
& x\in\mathcal X,\qquad y_e\in\mathcal Y_e(x),
\label{eq:hybrid_master_explicit}\\
& \xi_e\ge q_e^\top y_e-\eta,\qquad \xi_e\ge0,
\label{eq:hybrid_master_explicit_cvar}\\
& \xi_s\ge\theta_s-\eta,\qquad \xi_s\ge0,
\quad \forall s\in\mathcal D,
\label{eq:hybrid_master_decomp_cvar}\\
& \theta_s\ge a_{s\ell}^{\top}x+b_{s\ell},
\quad \forall \ell\in\mathcal K_s^{\mathrm{opt}},\,s\in\mathcal D,
\label{eq:hybrid_master_optcuts}\\
& f_{s\ell}^{\top}x+h_{s\ell}\le0,
\quad \forall \ell\in\mathcal K_s^{\mathrm{feas}},\,s\in\mathcal D ,
\label{eq:hybrid_master_feascuts}
\end{align}
\end{subequations}
where variables and parameters with subscript $e$ are linked to the embedded scenario. Here $\mathcal K_s^{\mathrm{opt}}$ and
$\mathcal K_s^{\mathrm{feas}}$ denote the retained optimality and
feasibility cuts. The former are obtained from dual-optimal SP solutions
and lower-approximate $Q_s(x)$, whereas the latter are obtained from
Farkas certificates and exclude first-stage points with infeasible
recourse.

\subsection{Learning Scenario Tail Relevance Online}

Solving \eqref{eq:hybrid_master} at iteration $k$ yields
$(x^k,\eta^k,\theta^k)$. Rather than evaluating every
$Q_s(x^k)$, OLAH-BD aims to identify the scenarios most likely to provide
useful Benders information. For an evaluated feasible scenario, define
\[
v_s^k=\left[Q_s(x^k)-\theta_s^k\right]_+
\]
and the CVaR-weighted residual
\begin{equation}
r_s^k=
\frac{\pi_s}{1-\alpha}\mathcal B_s^k v_s^k,
\qquad
\mathcal B_s^k=
\begin{cases}
1, & Q_s(x^k)>\eta^k,\\
0, & \text{otherwise}.
\end{cases}
\label{eq:risk_weighted_residual}
\end{equation}
Thus, $v_s^k$ measures master underestimation, while $r_s^k$ emphasizes
underestimated scenarios that also contribute to the current CVaR tail.

Since $Q_s(x^k)$ is unavailable for unsolved scenarios, the LR selector
uses four pre-solve features,
\begin{equation}
\phi_s^k=
\left[
\gamma(\widehat v_s^k),\,
\gamma(r_{s,\mathrm{last}}^k),\,
\tau_s^k,\,
\gamma(\bar r_s^k)
\right]^{\mathsf T},
\label{eq:lr_features}
\end{equation}
where $\gamma(z)=\ln(1+z)$ and
$\widehat v_s^k=[Q_s(x^{\kappa_s(k)})-\theta_s^k]_+$ uses the most
recent evaluation $\kappa_s(k)<k$ to estimate current master
underestimation. The remaining features record, respectively, the last
CVaR-weighted residual, an exponential moving average (EMA) of the threshold indicator
$\mathcal B_s^k$, and an EMA of the CVaR-weighted residual. The LR model assigns the usefulness score
\begin{equation}
p_s^k=
\sigma\!\left(w^\top\phi_s^k+b\right),
\qquad
\sigma(z)=\frac{1}{1+\exp(-z)},
\label{eq:lr_score}
\end{equation}
which ranks scenarios before SP evaluation. The model is trained on accumulated full-sweep data by minimizing a cross-entropy loss with \(\ell_2\) regularization.

Training labels are generated during full sweeps. Let
$\mathcal F_k\subseteq\mathcal D$ denote the set of decomposed SPs that
are feasible at iteration $k$, and define
\begin{equation}
m_{\alpha,k}=
\left\lceil(1-\alpha)|\mathcal F_k|\right\rceil .
\label{eq:tail_size}
\end{equation}
The tail-relevant set is then
\begin{equation}
\mathcal T_\alpha^k
=
\operatorname{Top}_{m_{\alpha,k}}
\left\{r_s^k:\,s\in\mathcal F_k\right\}.
\label{eq:tail_relevant_set}
\end{equation}
For the feasible scenarios, the training labels are
\begin{equation}
\ell_s^k=
\begin{cases}
1, & s\in\mathcal T_\alpha^k,\\
0, & s\in\mathcal F_k\setminus\mathcal T_\alpha^k,
\end{cases}
\qquad s\in\mathcal F_k .
\label{eq:training_label}
\end{equation}
The training history is updated as
\begin{equation}
\mathcal H\leftarrow
\mathcal H\cup
\left\{(\phi_s^k,\ell_s^k):s\in\mathcal F_k\right\},
\label{eq:training_set_update}
\end{equation}
so infeasible SPs are excluded from LR training and are handled
separately through the feasibility-cut mechanism.

\subsection{Adaptive Active Set, Cut Filtering, and Full Sweeps}

At adaptive iterations, only an active set
$\mathcal A_k\subset\mathcal D$ is evaluated:
\begin{equation}
\mathcal A_k=
\mathcal M_k
\cup\mathcal A_k^{\mathrm{LR}}
\cup\mathcal A_k^{\mathrm{stress}}
\cup\mathcal A_k^{\mathrm{stale}}
\cup\mathcal A_k^{\mathrm{rand}} ,
\label{eq:active_set}
\end{equation}
where $\mathcal M_k$ contains  infeasible scenarios from the previous iteration, while the remaining slots combine exploitation and safeguards. Approximately
$70\%$ of the current batch size are selected from the highest-ranked LR scenarios,
$10\%$ from high-stress scenarios \eqref{eq:stress_scenario}, $10\%$ from the stale scenario set, with
staleness defined as
$
a_s^k=k-\kappa_s(k),
$
and the remainder are sampled randomly. These components respectively
provide learned exploitation, physical-stress coverage, periodic
re-evaluation of neglected scenarios, and exploration against a
temporarily inaccurate LR model. 

An infeasible scenario contributes
its valid cut and enters $\mathcal M_{k+1}$. For feasible scenarios,
optimality-cut candidates are generated when
$Q_s(x^k)-\theta_s^k$ exceeds the prescribed tolerance and are filtered
according to CVaR relevance and residual magnitude. At full sweep iteration, all scenarios are evaluated and a valid upper bound is
obtained if all SPs are feasible, using
\begin{equation}
\overline z^k=
c^{\mathsf T}x^k+\min_\eta
\left[\eta+
\frac{1}{1-\alpha}
\sum_{s\in\mathcal S}
\pi_s
\left[Q_s(x^k)-\eta\right]_+\right] .
\label{eq:true_ub}
\end{equation}
Let $\widetilde{\mathcal A}_k$ denote the active set that the adaptive
policy selects during a full sweep. Selector quality is
measured against two metrics
\begin{equation}
R_{\mathrm{tail}}^k=
\frac{|\widetilde{\mathcal A}_k\cap\mathcal T_\alpha^k|}
{|\mathcal T_\alpha^k|},
\qquad
R_{\mathrm{risk}}^k=
\frac{\sum_{s\in\widetilde{\mathcal A}_k}r_s^k}
{\sum_{s\in\mathcal D}r_s^k},
\label{eq:selector_metrics}
\end{equation}
defined as tail recall, which measures how many truly tail-relevant scenarios are captured,
and risk coverage, which measures how much of the total CVaR-weighted
residual is represented, respectively. 

The first three iterations are full sweeps for initial cuts and LR
training, while adaptive selection starts from $k=4$ with a full sweep every
five iterations. The nominal batch size is selected as
\begin{equation}
K_0\approx2m_\alpha .
\label{eq:k_nominal}
\end{equation}
If a full sweep yields
$R_{\mathrm{tail}}^k<0.70$ or $R_{\mathrm{risk}}^k<0.70$, reveals
missed feasibility information, or substantial missed-cut activity, the
LR model is refreshed and
\begin{equation}
K_{k+1}
=
\min\left\{
K_{\max},
\max\left(K_k+10,\left\lceil1.5K_k\right\rceil\right)
\right\}.
\label{eq:k_validation_expand}
\end{equation}
If the infeasible fraction is at least $0.80$, a more aggressive
expansion is used,
\begin{equation}
K_{k+1}=\min\left\{K_{\max},2K_k\right\},
\label{eq:k_feas_expand}
\end{equation}
while two consecutive adaptive iterations with infeasible fraction at
most $0.30$ allow
\begin{equation}
K_{k+1}
=
\max\left\{K_0,\left\lceil0.7K_k\right\rceil\right\}.
\label{eq:k_shrink}
\end{equation}
Thus, OLAH-BD expands the computation effort when feasibility or LR model accuracy deteriorates and returns toward the nominal batch once the selector
stabilizes.
\section{Numerical Simulation}
\label{sec:numerical}

\subsection{Case Study and Implementation}

The proposed OLAH-BD method is evaluated on the BtM BESS sizing problem
defined in Appendix~\ref{app:bess_formulation}. The case study represents
a four-site residential energy community over a one-year hourly horizon.
The aggregate load is derived from RTE France data and scaled to an
annual consumption of 690.4 MWh, corresponding approximately to 100
households of mixed sizes. The load is distributed among the four sites
using fixed shares. PV generation is located at Site~1 and is obtained
for Marseille, France, using CAMS irradiance and Open-Meteo temperature
data. Joint PV--load uncertainty is represented using the VAR--KDE
scenario model described briefly in Appendix~\ref{app:scenario_generation}.

We consider two scenarios: S100, which uses 100 PV+load trajectories, and S500 which has 500 trajectories including those in S100. VBD is used
as the reference method, with all scenarios decomposed and evaluated at
every iteration. OLAH-BD instead embeds one stress-ranked scenario in
the RMP and applies the adaptive scenario-selection mechanism of
Section~\ref{sec:method} to the remaining scenarios. Both methods use
the same input data as well as solver settings to have a fair comparison. Simulation and algorithm settings are summarized in
Table~\ref{tab:case_params}. 

\begin{table}[t]
\caption{Case-study, algorithm, and solver parameters.}
\label{tab:case_params}
\centering
\scriptsize
\setlength{\tabcolsep}{3pt}
\renewcommand{\arraystretch}{1.04}
\begin{tabularx}{\columnwidth}{
>{\raggedright\arraybackslash}p{1.20cm}
>{\raggedright\arraybackslash}X}
\toprule
Type & Parameters \\
\midrule

Data
& 4 sites, $T=8760$ h, $\Delta t=1$ h, RTE France load,
690.4 MWh/year, site-wise load shares $[0.30,0.30,0.20,0.20]$ \\
& PV at Site~1: CAMS GHI + Open-Meteo temperature, area=2000 m$^2$,
VAR--KDE scenarios, uniform $\pi_s=1/|\mathcal S|$ \\
\midrule

BESS
& $c^E=40$ EUR/kWh-y, $c^P=20$ EUR/kW-y,
$\eta^{\rm ch}=0.95$, $\eta^{\rm dis}=0.96$,
$\rho^{\min/\max}=0.10/0.90$ \\
& $[\underline h,\overline h]=[1,10]$ h,
$c^{\rm deg}=0.005$ EUR/kWh throughput \\
\midrule

Grid/ACC
& $\lambda_t^{\rm buy}=0.20/0.15$ EUR/kWh (HP/HC),
$\lambda^{\rm exp}=0.05$ EUR/kWh \\
& $\tau_t^{\rm sh}$: TURPE MU4 ACC time-class tariff \cite{CRE2025TURPE7},
$c^{\rm sh}=12.12$ EUR/kW, $c^{\rm ret}=31.66$ EUR/kW,
$\overline G_i^{\rm sh}=\overline G_i^{\rm ret}=100$ kW \\
\midrule

OLAH-BD
& $\alpha=0.90$,
$(|\mathcal S|,|\mathcal D|)=(100,99)/(500,499)$,
$m_\alpha=10/50$,
$K_0=20/100$,
$K_{\max}=99/250$ \\
& One stress-ranked scenario embedded in the RMP,
3 initial full sweeps, certification every 5 iterations \\
& $\mathcal A_k$: 70\% LR, 10\% stale, 10\% stress, 10\% random \\
& Retrain if $R_{\rm tail}<0.70$ or $R_{\rm risk}<0.70$,
expand $\mathcal A_k$ if infeasible fraction $\ge0.80$,
shrink if $\le0.30$ for 2 adaptive iterations \\
\midrule

OLAH/VBD common
& Optimality/feasibility cut thresholds $=10^{-1}/10^{-4}$,
duplicate coefficient/RHS tolerances $=10^{-5}/10^{-6}$ \\
& Certified relative-gap tolerance $=10^{-4}$,
maximum 500 iterations \\
\midrule
System 
& Linux Mint 22.1, 13th Gen Intel Core i7-13800H
(14 cores, 20 threads), 32 GB RAM\\
\midrule
Gurobi v13.0.1
& RMP: concurrent- Dual Simplex (1 thread) + Barrier (3 threads), SP: dual simplex, 10 CPU threads for batchwise computation of SPs\\
\midrule
\end{tabularx}
\end{table}

\begin{table}[t]
\caption{Certified VBD and OLAH-BD results.}
\label{tab:main_results}
\centering
\scriptsize
\setlength{\tabcolsep}{2.6pt}
\renewcommand{\arraystretch}{1.02}
\begin{tabular}{lcc}
\toprule
Metric & S100 & S500 \\
\midrule
        & VBD/ OLAH-BD & VBD/ OLAH-BD\\
Iterations
& 75 / 25
& 69 / 35 \\

Certified UB (EUR)
& 77,535.25 / 77,534.01
& 78,836.45 / 78,836.84 \\

Certified gap (\%)
& 0.00197 / 0.00002
& 0.00026 / 0.00267 \\

SP evaluations
& 7,500 / 1,220
& 34,500 / 7,820 \\

Optimality cuts
& 4,759 / 823
& 27,203 / 5,112 \\

Feasibility cuts
& 1,991 / 27
& 4,027 / 0 \\

RMP time (s)
& 0.51 / 199.99
& 2.54 / 661.11 \\

SP time (s)
& 2,724.29 / 252.97
& 10,672.64 / 1,773.60 \\

Peak RAM (GB)
& 8.59 / 7.74
& 8.31 / 8.70 \\
\midrule
OLAH $E_1^\star$ (kWh)
& 65.85
& 56.59 \\

OLAH $P_1^\star$ (kW)
& 13.86
& 11.94 \\

No-BESS reference (EUR)
& 89,385.72
& 90,272.54 \\

Savings (\%)
& 13.26
& 12.67 \\
\bottomrule
\end{tabular}
\end{table}

\subsection{Results and Discussion}

\begin{figure}[t]
\centering
\subfloat[S100 bounds.]{
\includegraphics[width=0.47\columnwidth]
{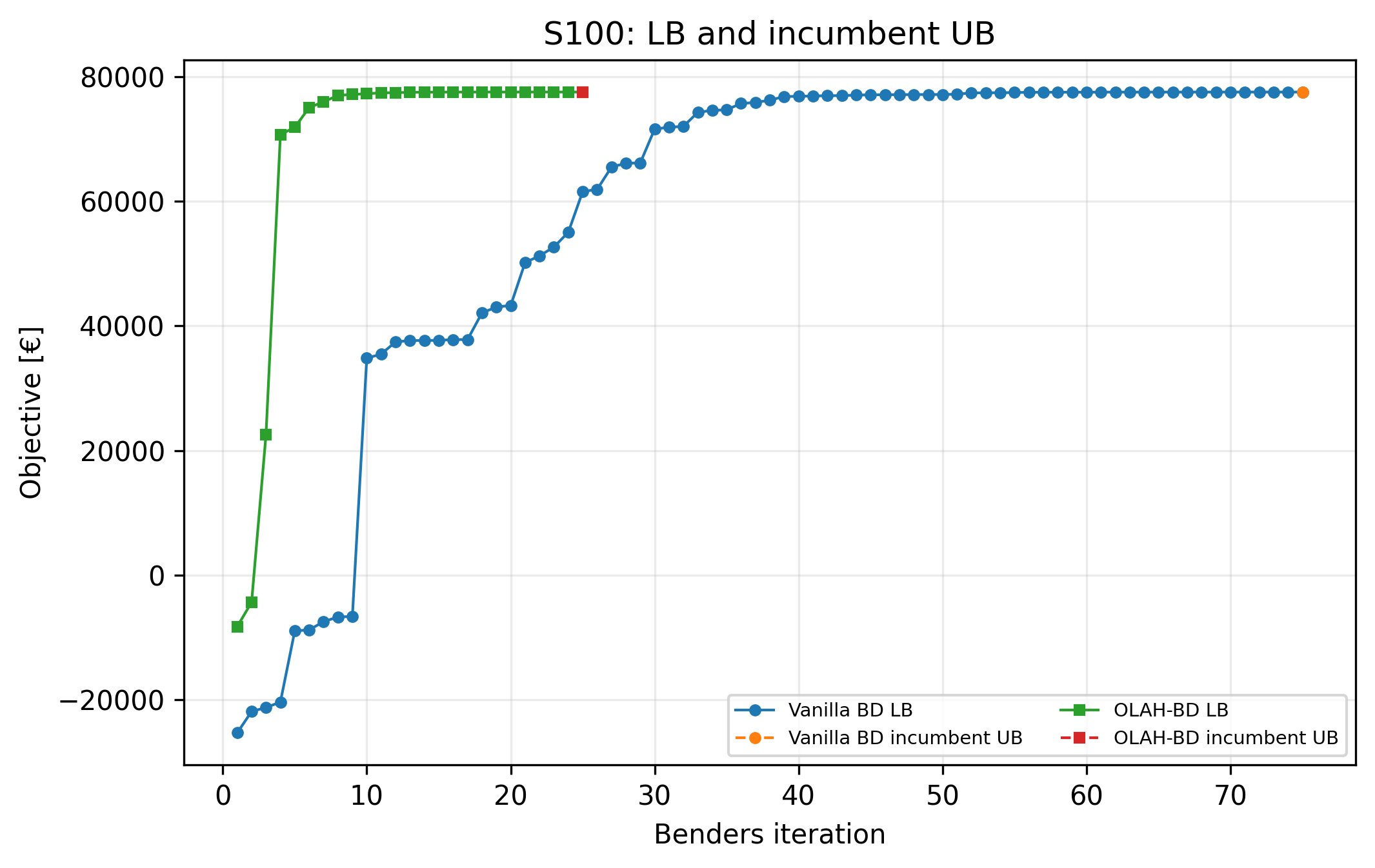}
\label{fig:bounds100}}
\hfill
\subfloat[S500 bounds.]{
\includegraphics[width=0.47\columnwidth]
{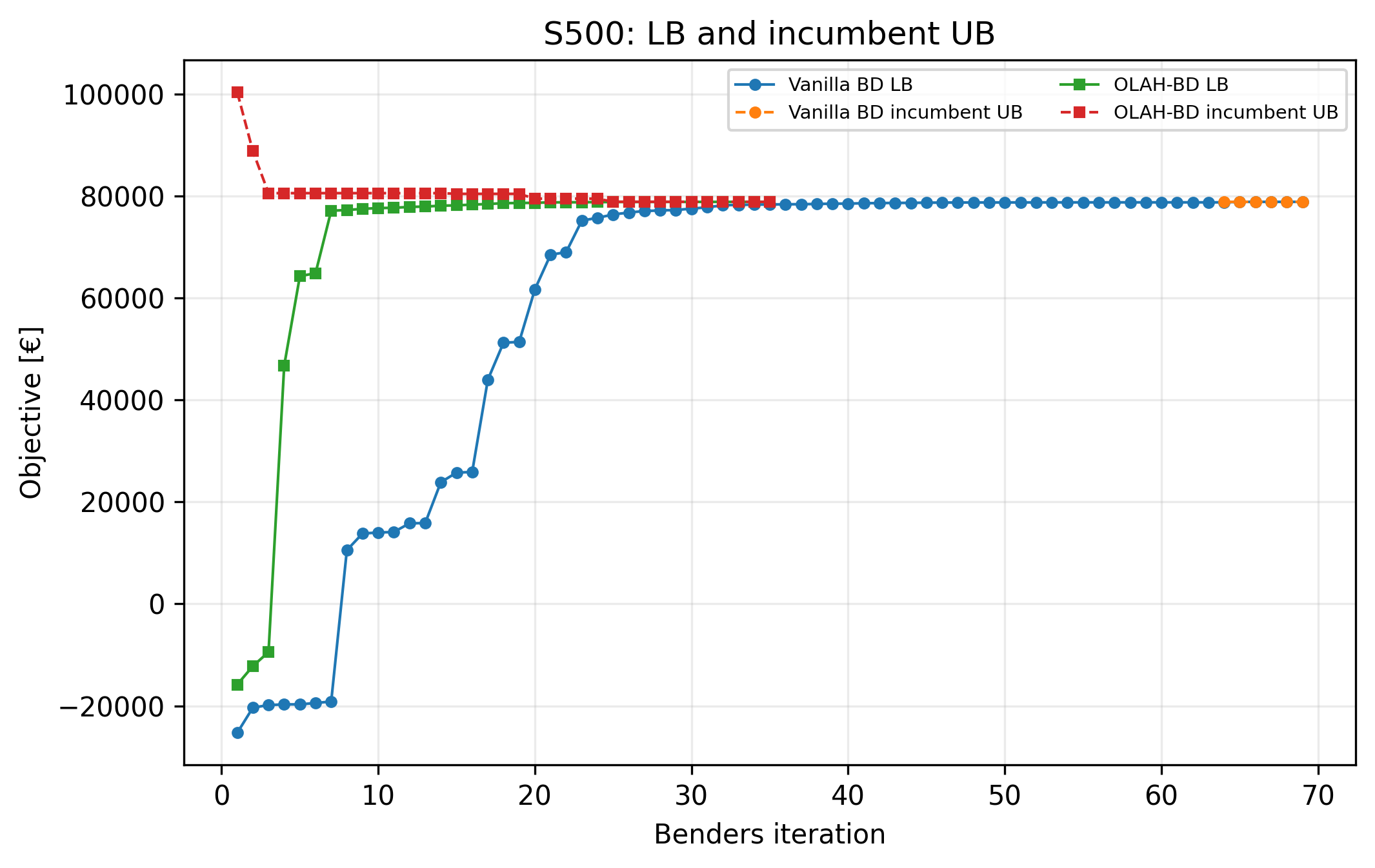}
\label{fig:bounds500}}

\vspace{-1mm}

\subfloat[S100 SP evaluations.]{
\includegraphics[width=0.47\columnwidth]
{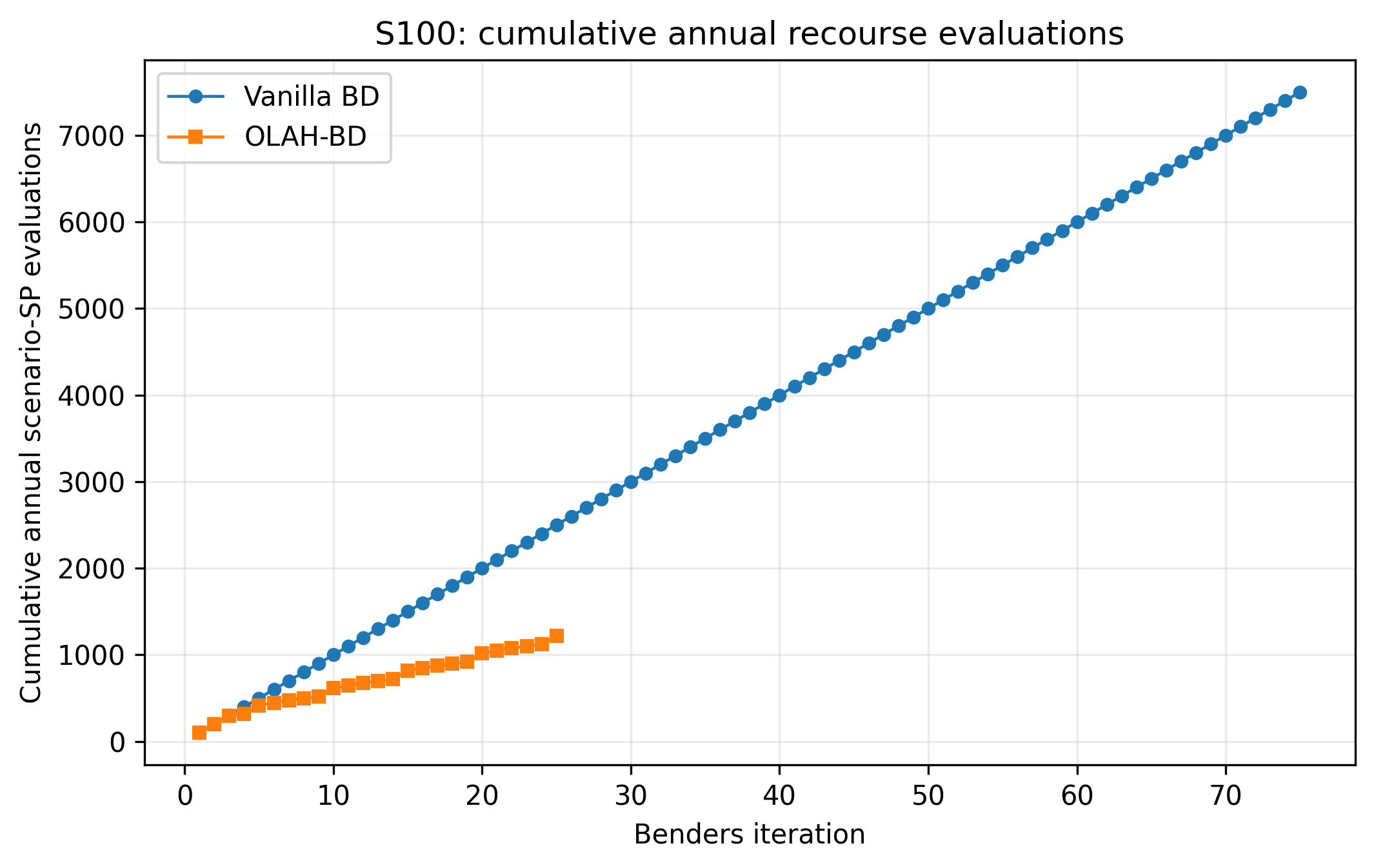}
\label{fig:sp100}}
\hfill
\subfloat[S500 SP evaluations.]{
\includegraphics[width=0.47\columnwidth]
{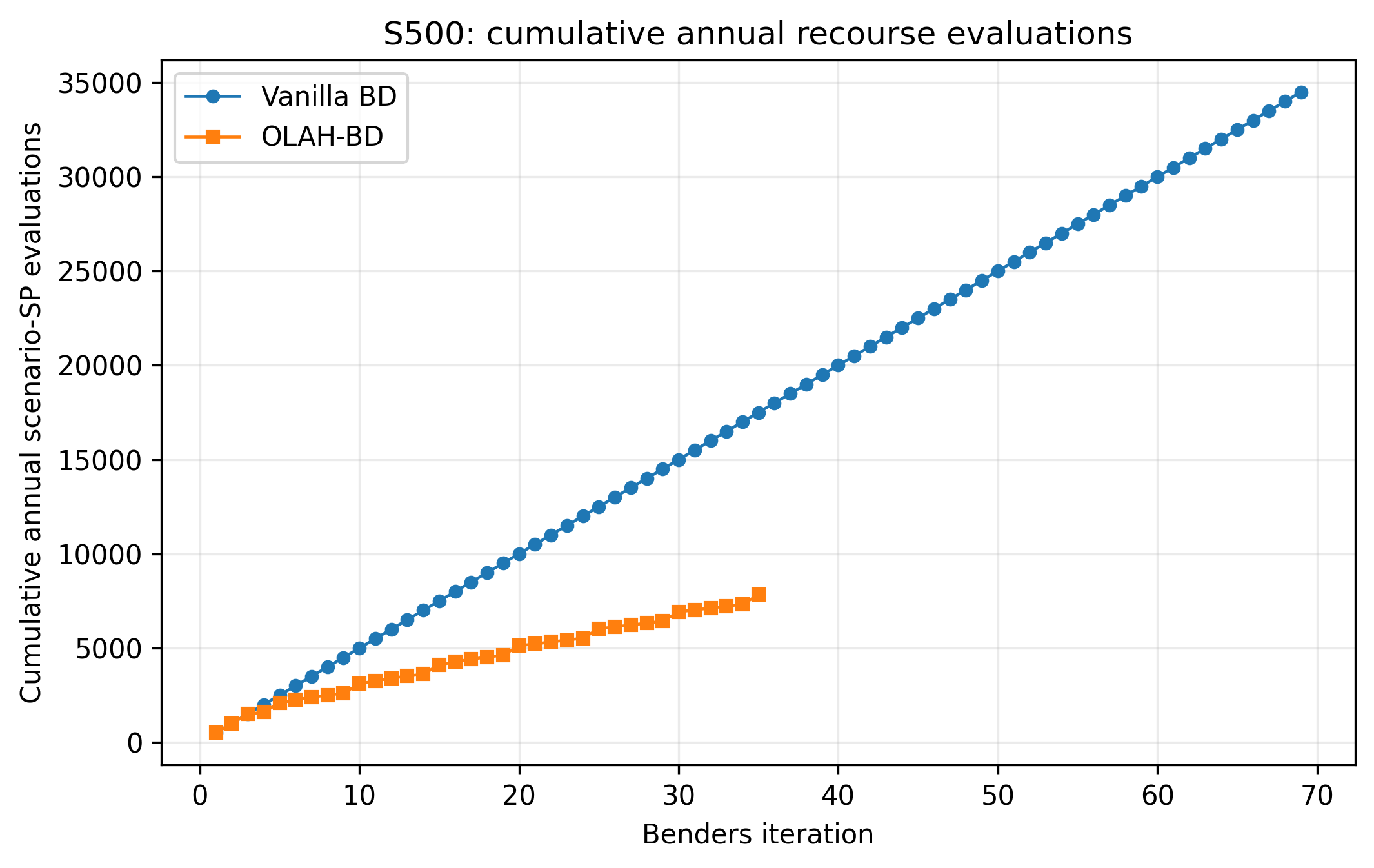}
\label{fig:sp500}}
\caption{Convergence and cumulative SP computations.}
\label{fig:performance_grid}
\end{figure}

Table~\ref{tab:main_results} compares the 
computational performance of VBD and OLAH-BD. The main computational advantage of OLAH-BD is the reduction in repeated
SP evaluation, which directly reflects in the total run time of the algorithm. While VBD evaluates the complete scenario set at every
iteration, OLAH-BD evaluates predominantly smaller adaptive batches (guided by the LR selector) and
uses full sweeps only for validation/certification. Consequently,
the reduction in SP workload dominates the additional cost of solving
the larger hybrid RMP and leads to a significantly lower total run
time i.e. nearly 6 times lower in S100 and about 4 times lower in S500. This pattern is clearly illustrated in Fig.~\ref{fig:performance_grid}, by the convergence and SP evaluation plots. Apart from the difference in per-iteration evaluations, the number of iterations for VBD is much higher than for OLAH-BD, which can be attributed to the hybrid RMP. In contrast, VBD progressively reconstructs this recourse information
through feasibility and optimality cuts, resulting in much slower LB improvement. However, embedding a single scenario does not guarantee feasibility for all scenarios and infeasible SPs can
still occur during adaptive iterations as is evidenced by the S100 run having 27 feasibility cuts. Instead, the objective is to reduce
the extent and persistence of early infeasibility rather than to enforce
complete recourse feasibility. 

The computational trade-off is also visible in the run-time and memory
statistics of Table~\ref{tab:main_results}. OLAH-BD spends more time in
the RMP primarily due to the embedded SP, but
this overhead is more than compensated by the reduction in SP computations, both through reduced iterations as well as fewer SP computations per iteration. Peak
memory remains of the same order for both methods, indicating that the
runtime improvement is obtained without a substantial change in the
overall memory requirement. This is to be expected, given that peak memory usage would occur during full sweeps in OLAH-BD and would roughly match that of VBD, which performs a full sweep every iteration, both methods use a maximum of 10 cpu threads in parallel for SP evaluations, while the RMP does a concurrent run (Dual Simplex on 1 thread + 3 threads for Barrier) on 4 cpu threads. It is worth noting that the memory footprint of OLAH-BD can be tuned to meet system specifications by adjusting the number of SP parallel threads and the use of RMP threads in the concurrent method. However, reducing the thread count might result in longer run times due to longer batch-wise computation of SPs and the Barrier method's sensitivity to the number of threads.

\begin{figure}[t]
\centering
\subfloat[S100 selector.]{
\includegraphics[width=0.47\columnwidth]
{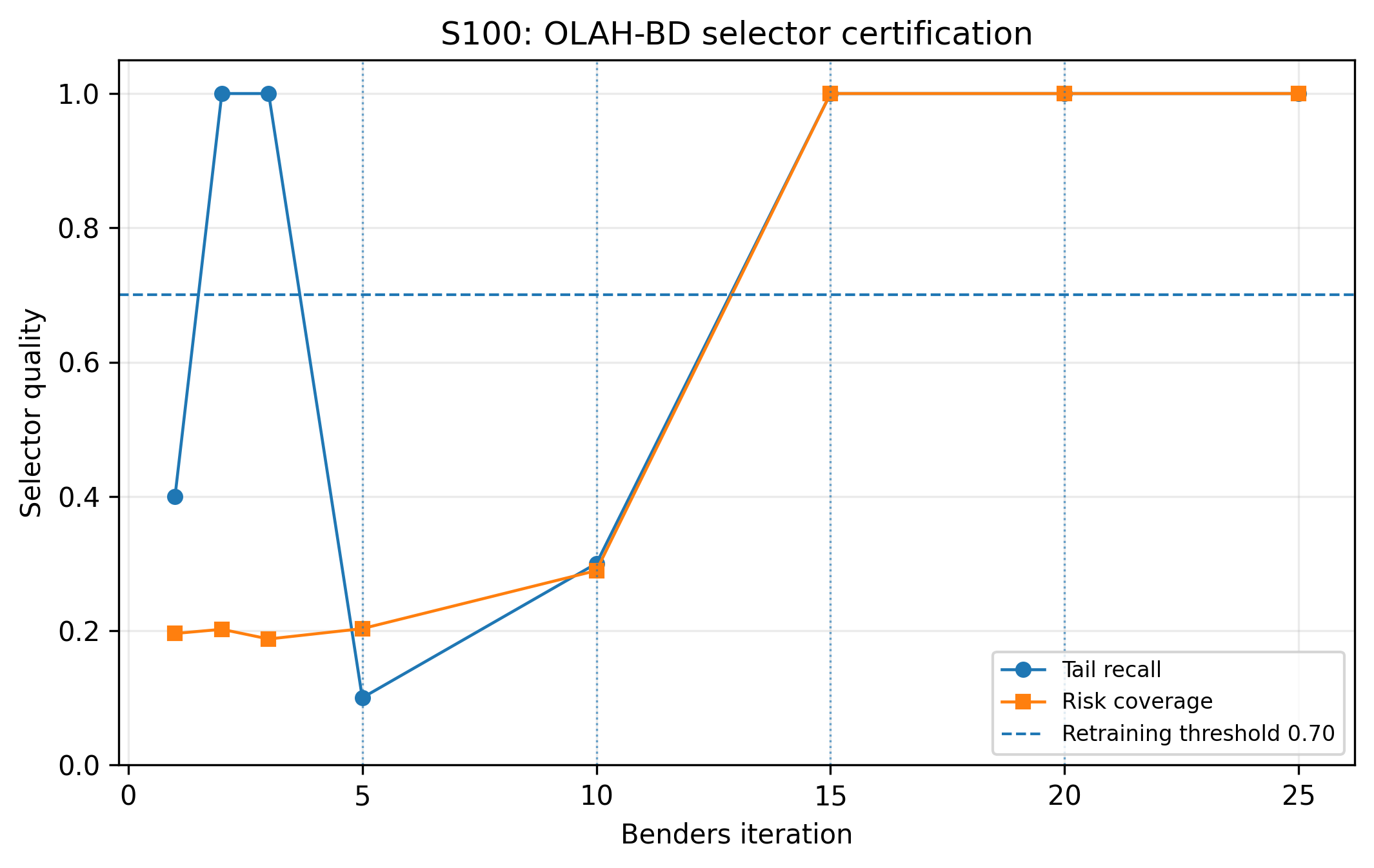}
\label{fig:tail100}}
\hfill
\subfloat[S500 selector.]{
\includegraphics[width=0.47\columnwidth]
{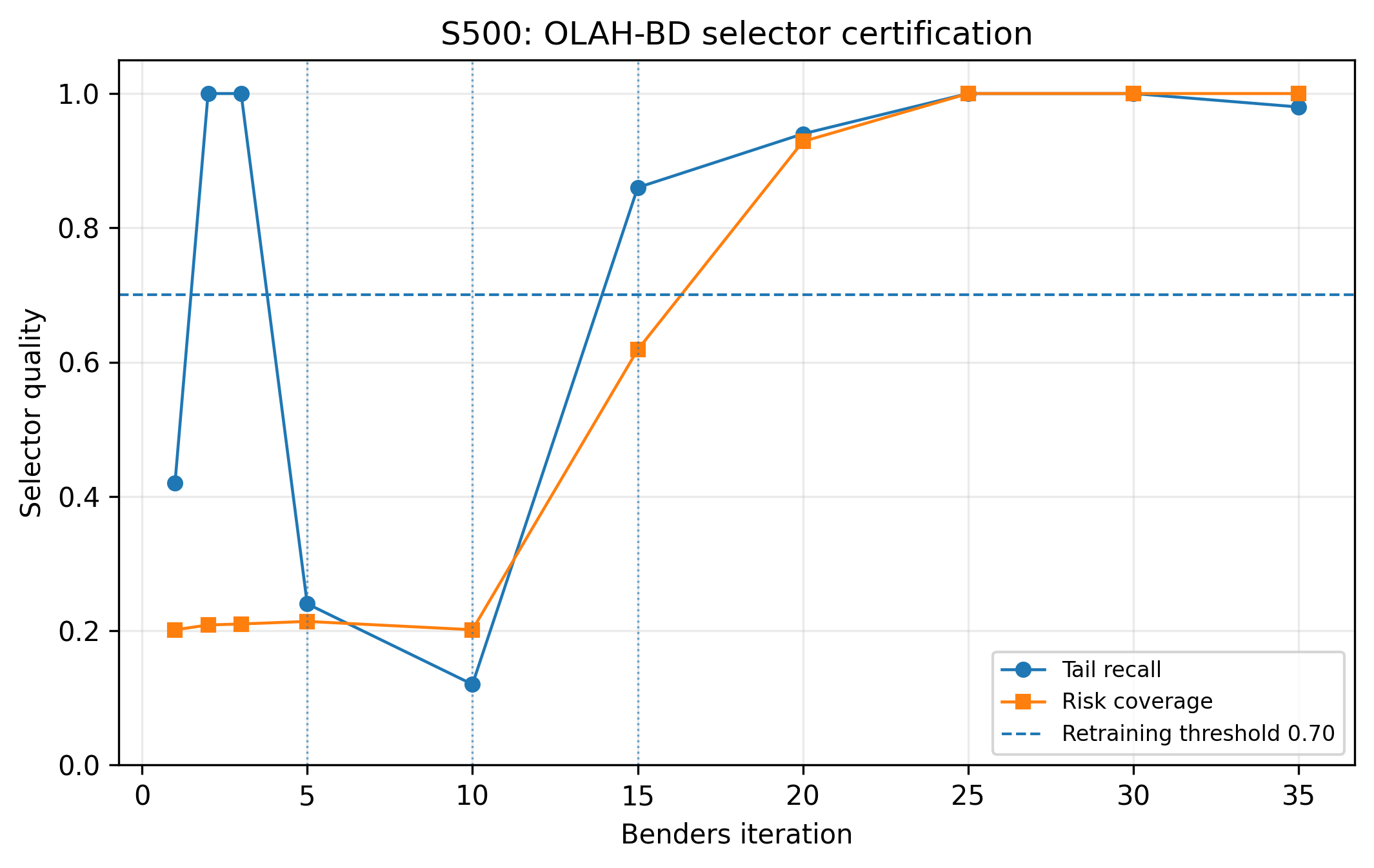}
\label{fig:tail500}}

\vspace{-1mm}

\subfloat[S100 batch control.]{
\includegraphics[width=0.47\columnwidth]
{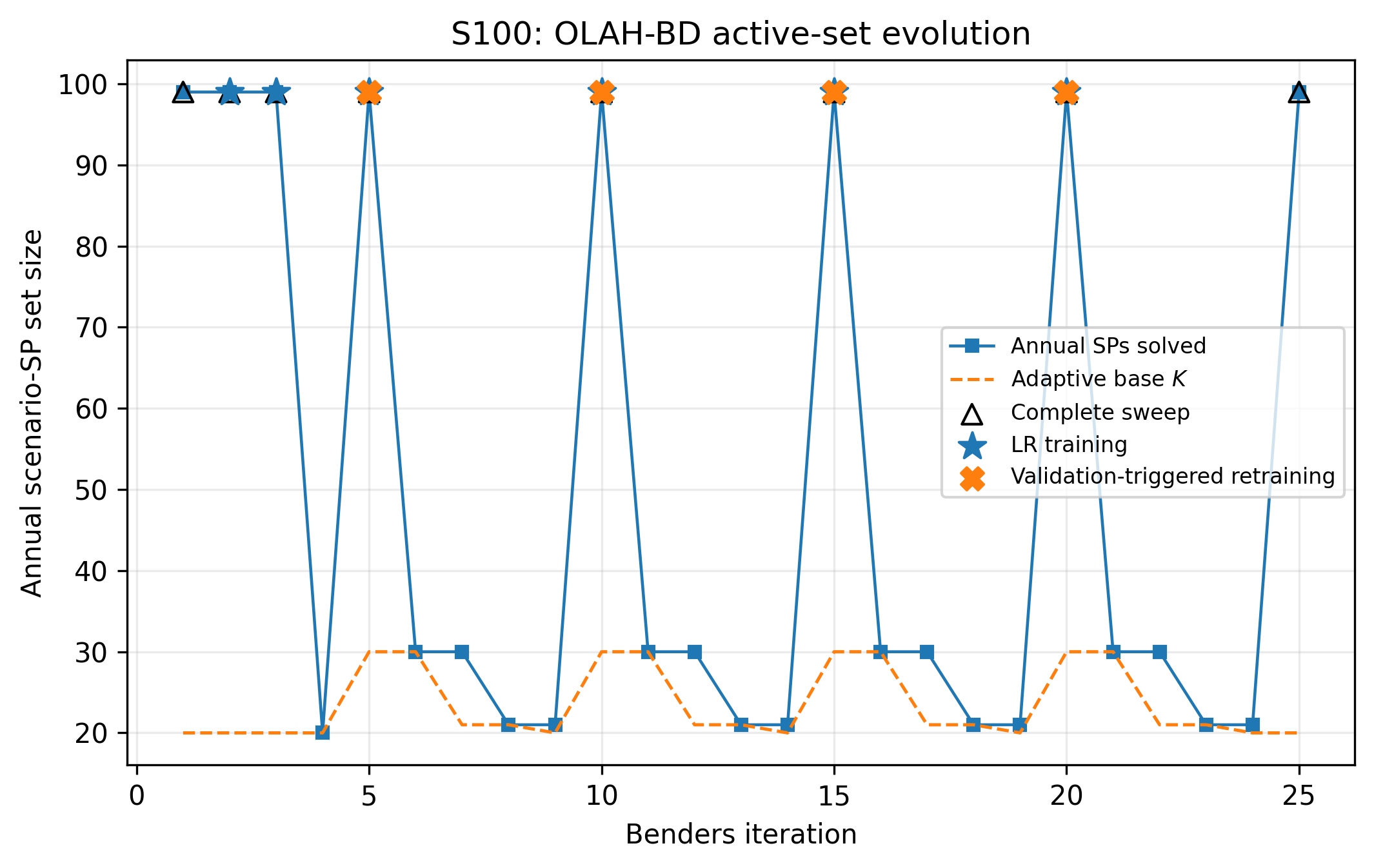}
\label{fig:batch100}}
\hfill
\subfloat[S500 batch control.]{
\includegraphics[width=0.47\columnwidth]
{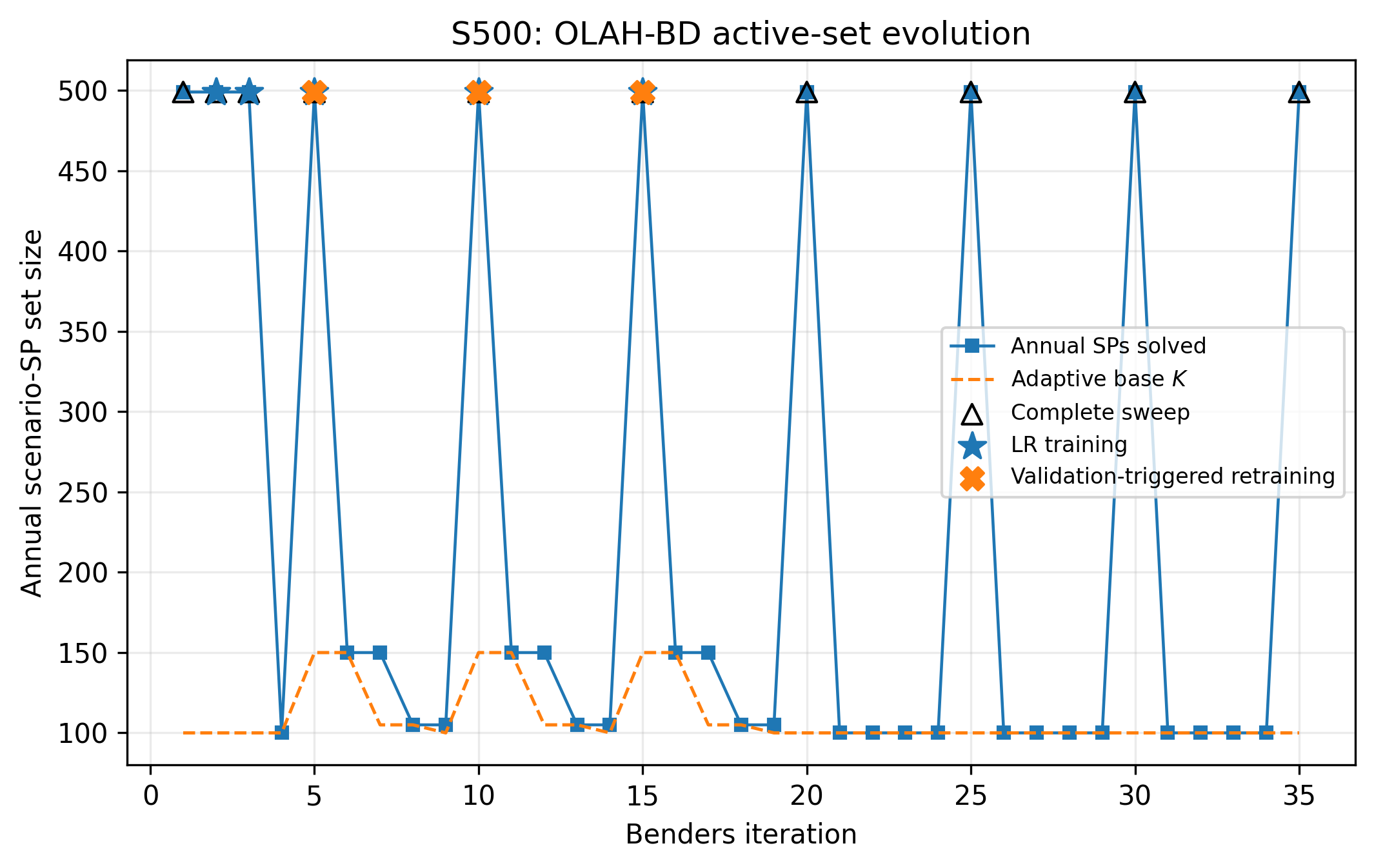}
\label{fig:batch500}}
\caption{LR SP selector validation and adaptive batch control.}
\label{fig:learning_grid}
\end{figure}

The LR selector behavior is illustrated in Fig.~\ref{fig:learning_grid}.
During the early iterations, the LR ranking is based on limited
training information and selector quality can vary. Periodic full sweeps
expose missed tail or feasibility information, trigger retraining when
required, and temporarily enlarge the active batch. As additional
full-sweep data become available, tail recall and risk coverage improve
and the batch returns toward its nominal size. The evolution pattern,
particularly in the case of S500, shows why scenario count alone
might be misleading.

\section{Conclusions and Outlook} \label{sec:conclusion}
This paper presents an accelerated online learning-based adaptive hybrid Benders approach for risk-averse optimal sizing of BtM BESS in energy communities. The main objective of this work is to demonstrate the capability of a lightweight ML model, such as logistic regression, to learn the significance of SPs in the CVaR tail from selected features, thereby enabling us to compute a small subset of SPs/iteration without compromising numerical accuracy. We illustrate that allowing the model to retrain during regime changes  improves its selection capabilities and that it scales well across scenarios despite the model's extremely small size. Current limitations of the work, such as the LP formulation of SPs and RMP, can potentially be addressed using different BD architectures, such as logic-based BD or column-and-constraint generation, within a robust optimization framework. It would be interesting to explore the performance of OLAH-BD with MILP RMP using callbacks and compare performance with the iterative approach. Another potential line of investigation could be to design and compare ML algorithms for SP selection using different feature sets.
\appendices

\section{Scenario Generation}
\label{app:scenario_generation}

The uncertainty set consists of joint PV-load trajectories. Historical PV and load data are decomposed into a deterministic calendar baseline and stochastic residuals. The PV baseline is computed by month and hour, while the load baseline is computed by month, day type, and hour. The residual vector is then modeled using a vector autoregressive process (VAR) with kernel density estimator (KDE) based sampled innovations \cite{saifiecon2023},
\begin{equation}
\epsilon_t
=
c+
\sum_{\ell=1}^{p}A_\ell\epsilon_{t-\ell}
+
u_t,
\qquad
u_t\sim \widehat f_u^{\mathrm{KDE}},
\label{eq:var_kde}
\end{equation}
where $\epsilon_t$ denotes the multivariate PV-load residual vector at time $t$, $c$ is the intercept vector, $A_\ell$ are the VAR coefficient matrices associated with lag $\ell$, and $p$ is the selected VAR order. The innovation term $u_t$ captures the residual stochastic variation and is sampled non-parametrically from the KDE $\widehat f_u^{\mathrm{KDE}}$. Separate VAR--KDE models are fitted by calendar groups. Simulated residual paths are added to the baseline and clipped to enforce physical limits, including nonnegative load, nonnegative PV, and zero PV during zero-baseline periods to avoid artificial night-time PV production. Details are omitted due to space limitations.

\section{Modeling Details for the Optimal Sizing Problem}
\label{app:bess_formulation}

For the considered behind-the-meter (BtM) BESS sizing problem, the
first-stage decision vector is
\[
x=
\left\{
E_i,\,
P_i,\,
\bar g_i^{\mathrm{sh}},\,
\bar g_i^{\mathrm{ret}}
\right\}_{i\in\mathcal I},
\]
where $E_i$ and $P_i$ are the BESS energy and power ratings, while
$\bar g_i^{\mathrm{sh}}$ and $\bar g_i^{\mathrm{ret}}$ denote the
subscribed capacities for community-shared and retailer energy,
respectively. The corresponding first-stage cost in
\eqref{eq:generic_cvar_2sp} is
\begin{equation}
c^\top x=
\sum_{i\in\mathcal I}
\left(
c^E E_i+c^P P_i+
c^{\mathrm{sh}}\bar g_i^{\mathrm{sh}}+
c^{\mathrm{ret}}\bar g_i^{\mathrm{ret}}
\right),
\label{eq:app_first_stage_cost}
\end{equation}
while the feasible set $\mathcal X$ is defined by
\begin{subequations}
\label{eq:app_first_stage_set}
\begin{align}
&0\le E_i\le\overline E_i,\qquad
0\le P_i\le\overline P_i,
&&\forall i\in\mathcal I,
\label{eq:app_storage_bounds}\\
&\underline hP_i\le E_i\le\overline hP_i,
&&\forall i\in\mathcal I,
\label{eq:app_duration_bounds}\\
&0\le\bar g_i^{\mathrm{sh}}\le\overline G_i^{\mathrm{sh}},
\qquad
0\le\bar g_i^{\mathrm{ret}}\le\overline G_i^{\mathrm{ret}},
&&\forall i\in\mathcal I .
\label{eq:app_subscription_bounds}
\end{align}
\end{subequations}

For a fixed first-stage decision $x$ and scenario $s$, $Q_s(x)$ is the
optimal annual operating cost. Let
$\mathcal T=\{1,\ldots,T\}$ denote the dispatch intervals and
$\mathcal T^E=\{1,\ldots,T+1\}$ the BESS energy-state indices.
The quantities $r_{its}$ and $d_{its}$ denote available PV generation
and demand, respectively. All recourse variables are nonnegative.

The notation $p_{its}^{a,b}$ denotes power transferred from source $a$
to destination $b$, where $\mathrm{pv}$, $\mathrm{bat}$,
$\mathrm{g}$, $\mathrm{sh}$, and $\mathrm{l}$ denote PV, BESS,
retailer grid, community shared-energy pool, and local load,
respectively. Thus, for example, $p^{\mathrm{pv,bat}}$ represents PV
charging of the BESS and $p^{\mathrm{bat,sh}}$ represents BESS export
to the shared-energy pool. PV may supply the local load, charge the
BESS, be shared within the community, exported to the grid, or curtailed,
the BESS may supply the local load or the shared-energy pool, while local
demand may additionally be supplied by shared-energy or retailer
imports. Unless otherwise stated, the following recourse constraints
hold for $i\in\mathcal I$ and $t\in\mathcal T$ for the fixed scenario
$s$.

The scenario recourse problem is
\begin{equation}
\label{eq:app_recourse}
\begin{split}
Q_s(x)=
\min_{y_s\ge0}\quad
&\sum_{i\in\mathcal I}\sum_{t\in\mathcal T}
\Big[
\lambda_t^{\mathrm{buy}}p_{its}^{\mathrm{g,l}}
+\tau_t^{\mathrm{sh}}
\left(
p_{its}^{\mathrm{sh,l}}
+p_{its}^{\mathrm{sh,bat}}
\right)
\\
&
+c^{\mathrm{deg}}
\left(
p_{its}^{\mathrm{pv,bat}}
+p_{its}^{\mathrm{sh,bat}}
+p_{its}^{\mathrm{bat,l}}
+p_{its}^{\mathrm{bat,sh}}
\right)\\
&
-\lambda^{\mathrm{exp}}p_{its}^{\mathrm{pv,g}}
\Big]\Delta t .
\end{split}
\end{equation}

PV generation is allocated according to
\begin{equation}
p_{its}^{\mathrm{pv,bat}}
+p_{its}^{\mathrm{pv,l}}
+p_{its}^{\mathrm{pv,g}}
+p_{its}^{\mathrm{pv,sh}}
+p_{its}^{\mathrm{curt}}
=
r_{its}.
\label{eq:app_pv_balance}
\end{equation}
Local demand is supplied by PV, BESS discharge, community-shared energy,
or retailer imports:
\begin{equation}
p_{its}^{\mathrm{bat,l}}
+p_{its}^{\mathrm{pv,l}}
+p_{its}^{\mathrm{sh,l}}
+p_{its}^{\mathrm{g,l}}
=
d_{its}.
\label{eq:app_load_balance}
\end{equation}

The BESS energy dynamics are
\begin{equation}
\begin{split}
e_{i,t+1,s}
=&\,
e_{its}
+\eta^{\mathrm{ch}}\Delta t
\left(
p_{its}^{\mathrm{pv,bat}}
+p_{its}^{\mathrm{sh,bat}}
\right)
\\
&-\frac{\Delta t}{\eta^{\mathrm{dis}}}
\left(
p_{its}^{\mathrm{bat,l}}
+p_{its}^{\mathrm{bat,sh}}
\right),
\end{split}
\label{eq:app_soc_dynamics}
\end{equation}
with initial and cyclic terminal conditions
\begin{equation}
e_{i1s}=\rho_i^0E_i,
\qquad
e_{i,T+1,s}=e_{i1s},
\qquad
\forall i\in\mathcal I .
\label{eq:app_soc_initial_final}
\end{equation}
The admissible energy range is
\begin{equation}
\rho^{\min}E_i
\le e_{its}\le
\rho^{\max}E_i,
\qquad
\forall i\in\mathcal I,\,
t\in\mathcal T^E .
\label{eq:app_soc_limits}
\end{equation}
The aggregate BESS charging and discharging powers are bounded by the
first-stage power rating:
\begin{equation}
p_{its}^{\mathrm{pv,bat}}
+p_{its}^{\mathrm{sh,bat}}
\le P_i,
\label{eq:app_charge_limit}
\end{equation}
\begin{equation}
p_{its}^{\mathrm{bat,l}}
+p_{its}^{\mathrm{bat,sh}}
\le P_i.
\label{eq:app_discharge_limit}
\end{equation}

At each time step, total exports to the community-shared pool must equal
total withdrawals:
\begin{equation}
\sum_{i\in\mathcal I}
\left(
p_{its}^{\mathrm{pv,sh}}
+p_{its}^{\mathrm{bat,sh}}
\right)
=
\sum_{i\in\mathcal I}
\left(
p_{its}^{\mathrm{sh,l}}
+p_{its}^{\mathrm{sh,bat}}
\right),
\qquad
\forall t\in\mathcal T .
\label{eq:app_community_balance}
\end{equation}

Shared-energy imports and exports are limited by the optimized
shared-energy subscription,
\begin{equation}
p_{its}^{\mathrm{sh,l}}
+p_{its}^{\mathrm{sh,bat}}
\le\bar g_i^{\mathrm{sh}},
\label{eq:app_shared_import_sub}
\end{equation}
\begin{equation}
p_{its}^{\mathrm{pv,sh}}
+p_{its}^{\mathrm{bat,sh}}
\le\bar g_i^{\mathrm{sh}},
\label{eq:app_shared_export_sub}
\end{equation}
while retailer imports satisfy
\begin{equation}
p_{its}^{\mathrm{g,l}}
\le\bar g_i^{\mathrm{ret}}.
\label{eq:app_retail_sub}
\end{equation}

Finally, the total simultaneous import and export flows are constrained
by the fixed physical point-of-connection capacity
$\overline G_i$:
\begin{equation}
p_{its}^{\mathrm{g,l}}
+p_{its}^{\mathrm{sh,l}}
+p_{its}^{\mathrm{sh,bat}}
\le\overline G_i,
\label{eq:app_poc_import}
\end{equation}
\begin{equation}
p_{its}^{\mathrm{pv,g}}
+p_{its}^{\mathrm{pv,sh}}
+p_{its}^{\mathrm{bat,sh}}
\le\overline G_i.
\label{eq:app_poc_export}
\end{equation}

Constraints \eqref{eq:app_pv_balance}--\eqref{eq:app_load_balance}
allocate PV generation and satisfy local demand.
Equations \eqref{eq:app_soc_dynamics}--\eqref{eq:app_discharge_limit}
describe BESS operation and couple scenario dispatch to the first-stage
energy and power ratings. Equation \eqref{eq:app_community_balance}
balances the shared-energy pool, while
\eqref{eq:app_shared_import_sub}--\eqref{eq:app_retail_sub} enforce the
optimized shared-energy and retailer subscriptions. Finally,
\eqref{eq:app_poc_import}--\eqref{eq:app_poc_export} impose the physical
import/export rating of each site's grid connection.

\bibliographystyle{IEEEtran}
\bibliography{new_biblio}   

\end{document}